\documentclass[12pt,a4paper,fleqn]{article}
\usepackage[T1]{fontenc}
\usepackage{amsmath}
\usepackage{amssymb} % e.g. \gtrsim
\usepackage{bbm} % BlackBoeard letters
\usepackage[small]{caption2} % font of captions is different from text
\usepackage{graphicx} % \including PostScript
\usepackage{mathrsfs}	% e.g. nice Lagrange-L with \mathscr{L}
\usepackage[small,loose]{subfigure}  % subfigures with (a), (b) etc., ... and own caption
\usepackage{cite} % [1-5] instead of [1,2,3,4,5]	
\usepackage{cancel} 
\usepackage{color} 
\newcommand{\I}{\mathrm{i}}

\newcommand{\eVdist}{\kern-0.06em}
\newcommand{\gev}{\:\text{Ge\eVdist V}}

\newcommand{\higgs}{\ensuremath{{h_1}}}
\newcommand{\singlino}{\ensuremath{{\widetilde{s}}}}
\newcommand{\pseudoscalar}{\ensuremath{a_s}}

\DeclareMathAlphabet{\mathpzc}{OT1}{pzc}{m}{it}

\allowdisplaybreaks
\begin{document}
\thispagestyle{empty}
\begin{titlepage}

\begin{flushright}
TUM-HEP 772/10\\
MPP-2010-130
\end{flushright}

\vspace*{1.0cm}

\begin{center}
\Huge\textbf{Light dark matter in the singlet-extended MSSM}
\end{center}
\vspace{1cm}
 \center{
\textbf{
Rolf Kappl\footnote[1]{Email: \texttt{rolf.kappl@ph.tum.de}}$^{,ab}${}, 
Michael Ratz\footnote[2]{Email: \texttt{mratz@ph.tum.de}}$^{,a}$, 
Martin Wolfgang Winkler\footnote[3]{Email: \texttt{mwinkler@ph.tum.de}}$^{,a}$
}
}
\\[5mm]
\begin{center}
\textit{$^a$\small
~Physik-Department T30, Technische Universit\"at M\"unchen, \\
James-Franck-Stra\ss e, 85748 Garching, Germany
}
\\[5mm]
\textit{$^b$\small
~Max-Planck-Institut f\"ur Physik (Werner-Heisenberg-Institut),\\
F\"ohringer Ring 6,
80805 M\"unchen, Germany
}
\end{center}

\date{}
\vspace{1cm}

\begin{abstract}
We discuss the possibility of light dark matter in a general singlet extension
of the MSSM. Singlino LSPs with masses of a few GeV can explain the signals
reported by the CRESST, CoGeNT and possibly also DAMA experiments.
The interactions between singlinos and nuclei are mediated by a
scalar whose properties coincide with those of the SM Higgs up to two crucial
differences: the scalar has a mass of a few GeV and its interaction strengths are
suppressed by a universal factor. We show that such a scalar can be consistent
with current experimental constraints, and that annihilation of singlinos into
such scalars in the early universe can naturally lead to a relic abundance
consistent with the observed density of cold dark matter.
\end{abstract}

\end{titlepage}

\newpage

\section{Introduction}

Supersymmetry offers a very attractive solution to the dark matter puzzle since
the lightest superpartner (LSP) is a quite compelling candidate for the observed
cold dark matter (CDM). The LSP is usually assumed to be stable or, at least,
long lived. Most studies on scenarios of supersymmetric dark matter focus on the
LSPs with electroweak scale masses and cross sections, whose relic density can
match the measured CDM density. However, recent results from the
direct detection experiments CoGeNT~\cite{Aalseth:2010vx} and
CRESST~\cite{Seidel:2010nn} seem to hint at somewhat lighter dark matter
particles with masses of a few $\mathrm{GeV}$. This interpretation is also
consistent with the DAMA signal~\cite{Bernabei:2008yi,Bernabei:2010mq}. Do we
expect to have such particles in supersymmetric extensions of the standard model
(SM)?  Certainly, in the minimal supersymmetric SM (MSSM), such masses appear
hardly justifiable for particles interacting strong enough to explain the above
signals~\cite{Hooper:2002nq,Bottino:2002ry,Feldman:2010ke}. This is because, in
the MSSM, such particles will typically contribute to the $Z$ boson decay width.
On the other hand, in singlet extensions of the MSSM this
problem may be circumvented.  While in the usual NMSSM it still appears
difficult~\cite{Das:2010ww,Hooper:2009gm}, but not
impossible~\cite{Draper:2010ew}, to obtain particles with the
desired properties, generalized singlet extensions of the
MSSM~\cite{Belikov:2010yi} can indeed give rise to settings with light dark
matter candidates whose interactions with nuclei are mediated by weakly coupled
light scalars.

In this paper we focus on a particularly simple scenario based on such a singlet
extension of the MSSM in which the singlet sector is only weakly coupled to the
MSSM. As we shall see, this will lead to a Higgs-like scalar $h_1$ which behaves
similar to the SM Higgs with two crucial differences. First, all couplings to SM
matter are suppressed by an overall factor, and second its mass is in the GeV
range. This scalar is accompanied by a singlino superpartner, which, as we shall
see, has naturally the correct abundance to explain the observed CDM. Further,
$h_1$ mediated scatterings of nuclei with the singlino can give rise to the
signals reported by CRESST, CoGeNT and possibly also DAMA.

\section{Light Singlets in the S-MSSM}

We consider the MSSM Higgs sector extended by a gauge singlet superfield $S$.
The most general renormalizable superpotential
reads~\cite{Drees:1988fc}\footnote{A possible linear term in $S$ can be absorbed
into the quadratic and cubic terms \cite{Drees:1988fc}.} 
\begin{equation}
 \mathscr{W}~ =~ \mu\, H_u H_d + \lambda\, S\, H_u H_d + 
 \frac{\mu_s}{2}\, S^2 + \frac{\kappa}{3}\, S^3\;.
\end{equation}
In the so-called NMSSM the $\mu$ and $\mu_s$ terms are set to zero.  On the
other hand, it is well-known that there are mechanisms that explain a suppressed
$\mu$ term \cite{Kim:1983dt,Giudice:1988yz}; such mechanisms may also give rise
to a $\mu_s$ parameter of the order of the electroweak scale. Recently the
resulting scheme has been investigated in a different context and was dubbed
`S-MSSM' \cite{Delgado:2010cw}; we will adopt this terminology.
A simple setting where the smallness of $\mu$ and $\mu_s$ finds an explanation
will be discussed elsewhere~\cite{LRRRSSV2}. We include both dimensionful
parameters in our analysis. In addition the scalar potential includes the
following soft terms
\begin{eqnarray}
 V_{\text{soft}}&=& m_{h_u}^2\, |h_u|^2
 + m_{h_d}^2\, |h_d|^2 
 + m_s^2\, |s|^2 \nonumber \\
&&
{}+\left(   B\mu\, h_u h_d 
 + \lambda\, A_{\lambda}\, s\, h_u h_d 
 + \frac{B\mu_s}{2}\, s^2 + \frac{\kappa}{3}\, A_\kappa s^3 + \text{h.c.}\right) \;.
\end{eqnarray}
The singlet superfield $S$ contains a complex scalar $s$ and a Majorana fermion, the singlino $\singlino$. 

The important feature of the resulting model is that all interactions between
the MSSM and the singlet sectors are controlled by a single parameter $\lambda$.
As we shall see, in the region where $\lambda$ is of the order $10^{-2\dots3}$
and the singlet fields are light a simple explanation of the
direct detection signals mentioned in the introduction emerges.
To obtain light singlets we shall assume that all singlet mass
terms are set by a scale $m_\text{singlet}\sim 10\gev$. Relatively suppressed
soft terms for the singlet can be motivated in settings in which the MSSM soft
masses are dominated by the gaugino contribution in the renormalization group.
In what follows, we start by discussing the limit $\lambda\to0$, and then
explore what happens if we switch on a finite but small $\lambda$.

\paragraph{Limit $\boldsymbol{\lambda\to0}$.}
All terms which mix the singlets with the MSSM contain the parameter $\lambda$,
i.e.\ in the case $\lambda=0$ both sectors are completely decoupled. The singlino
mass is simply given by $m_{\singlino}=\mu_s$, the complex scalar $s$ receives
additional mass contributions from the soft terms which also split its real and
imaginary components. If we introduce the real scalar $h_s$ and pseudoscalar
$\pseudoscalar$ through the relation $s=(h_s + \I\,\pseudoscalar)/\sqrt{2}$, we find
$m_{h_s}^2=m_s^2+\mu_s^2+B\mu_s$ and $m_{\pseudoscalar}^2=m_s^2+\mu_s^2-B\mu_s$.
A light singlet sector can be obtained if we assume that all mass parameters
$m_s^2,\mu_s^2,B\mu_s\sim m_{\text{singlet}}^2$ with $m_{\text{singlet}}=\mathcal{O}(10\gev)$. The following
discussion is based on this assumption.

\paragraph{Small $\boldsymbol{\lambda}$.}
Switching on a small $\lambda$ leads to couplings to and mixings with the MSSM
fields.
Through the $F$- and soft terms of the MSSM Higgs fields there arises a linear term in $s$
of the form $\lambda\, \mu_\text{eff}\, v_\mathrm{EW}^2\, s$, where we introduced $\mu_\text{eff}=\mu - v_1 v_2 A_\lambda/v_\mathrm{EW}^2$. Here $v_1=\langle h_d \rangle$, $v_2=\langle h_u \rangle$ and $v_\mathrm{EW}^2=v_1^2+v_2^2\simeq (174\gev)^2$. The linear term induces a vacuum expectation value
$x=\langle s \rangle$ which can be estimated as
\begin{equation}
 x~\sim~\lambda\, \frac{v_\mathrm{EW}^2}{m_\text{singlet}^2}\,\mu_\text{eff}\;.
\end{equation}
There are two competing effects, the smallness of $\lambda$ and the
$m_\text{singlet}^2$ in the denominator, such that $x$ can be of the order of
the electroweak scale. Note, however, that the impact of $x$ on the SM Higgs
masses is almost negligible.
In the presence of the singlet VEV there will be new singlet mass terms such as
$\kappa^2\,s^2$ and $\kappa\,A_\kappa\,s$. Therefore, in order to keep the
singlet sector light, we assume that the self-coupling $\kappa$ is not too large,
$\kappa\lesssim 0.1$, and that the trilinear coupling $A_\kappa \lesssim
m_\text{singlet}$.

Let us look at the masses and mixings of the singlets. As $\lambda$ is small we
can treat MSSM and singlet sector separately and consider mixing as a
perturbation. To simplify our analysis, we impose the decoupling limit on the
MSSM Higgs fields. This allows us to ignore mixing of the singlets with the
pseudoscalar and the heavy scalar MSSM Higgs. We, however, keep track of the
mixing between the light MSSM Higgs $h$ and $h_s$. We further use the
minimization conditions for the Higgs potential in order to eliminate the soft
masses.

The mass of the singlet pseudoscalar is then given by
\begin{equation}
 m_{\pseudoscalar}^2~\simeq~-2 B\mu_s - x\, \kappa \, (3 A_\kappa+\mu_s)
 -\lambda\,\frac{\mu_\text{eff}}{x}\, v_\mathrm{EW}^2\;.
\end{equation}
The scalar mass matrix in the basis $(h,h_s)$ reads
\begin{equation}
 M_H^2 ~=~ \begin{pmatrix}
          m_h^2 & m_{h h_s}^2 \\  
          m_{h h_s}^2 & m_{h_s}^2
         \end{pmatrix}
\end{equation}
with
\begin{eqnarray}
 m_{h_s}^2 &\simeq& \kappa\, x\, (A_\kappa +4 \kappa x + 3 \mu_s)
 -\lambda\,\frac{\mu_\text{eff}}{x}\,v_\mathrm{EW}^2
 \;,\\
 m_{h h_s}^2 
 &\simeq& 
 2\lambda\, v_\mathrm{EW}\, \mu_{\text{eff}}\;,
\end{eqnarray}
and $m_h^2$ as in the usual MSSM.
Note that with the assumptions made all contribution to $m_{\pseudoscalar}$ and $m_{h_s}$
are of the order $m_\text{singlet}$ or smaller, i.e.\ we obtain $m_{\pseudoscalar},m_{h_s}
\sim m_\text{singlet}$. Given our assumptions, $m_{h h_s} \sim
m_\text{singlet}$.

As $m_h^2 \gg m_{h h_s}^2, \; m_{h_s}^2$ there is little mixing between $h$ and
$h_s$. The light physical mass eigenstate is mainly singlet with a small admixture from
$h$,
\begin{equation}
 \higgs ~\simeq~ \cos\theta \,h_s - \sin\theta \,h
\end{equation}
with
\begin{eqnarray}
 \cos\theta &\simeq& 1\;,\\
 \sin\theta &\simeq& \frac{m_{h h_s}^2}{m_h^2}\;.
\end{eqnarray}
The heavier state $h_2$ essentially coincides with the MSSM Higgs $h$.
The mass of $\higgs$ is given by
\begin{equation}
 m_\higgs^2 ~\simeq~ m_{h_s}^2 - \frac{m_{h h_s}^4}{m_h^2} \;.
\end{equation}

In the fermion sector there is mixing between the singlino and the MSSM
neutralinos. This mixing can maximally reach the size of $\sin\theta$ if the
higgsinos are relatively light.\footnote{The possible exception
of a bino with a mass close to $m_\singlino$ is not considered here.} As such
a small mixing in the fermion sector does not play a role in the following
discussion, we ignore it and take the LSP to be a pure singlino with mass
\begin{equation}
 m_{\singlino}~=~\mu_s + 2 \kappa\, x\;.
\end{equation}

The couplings in the singlet sector are all controlled by $\kappa$. Most
relevant for the following discussion are the trilinear interaction terms which
comprise
\begin{equation}
 \mathscr{L} ~\supset~ -\frac{1}{2} \,g_{\higgs \singlino\singlino}\, 
 \higgs \,\overline{\singlino}\,\singlino 
 -\frac{1}{2}\,g_{\pseudoscalar \singlino\singlino}\, \pseudoscalar \,\overline{\singlino} \gamma_5\singlino -\frac{1}{6} \,g_{\higgs \higgs \higgs} \,\higgs^3 
 - \frac{1}{2}\, g_{\higgs \pseudoscalar \pseudoscalar}\, \higgs\, \pseudoscalar^2
\end{equation}
with
\begin{subequations}\label{eq:couplings}
\begin{eqnarray}
 g_{\higgs \singlino\singlino} &\simeq& \sqrt{2} \kappa\;, \\
 g_{\pseudoscalar \singlino\singlino} &\simeq& -\I\, \sqrt{2}\, 
 \kappa\;,\\
 g_{\higgs \higgs \higgs} &\simeq& \sqrt{2} \kappa \,(3\,m_{\singlino} +
 A_\kappa)\;, \\ 
 g_{\higgs \pseudoscalar \pseudoscalar} &\simeq& \sqrt{2} \kappa\, (m_{\singlino}
 - A_\kappa)\;.
\end{eqnarray}
\end{subequations}
The coupling of $\higgs$ to quarks and leptons is the SM Higgs
coupling suppressed by a factor of $\sin{\theta}$.

In summary we have obtained a light scalar which shares the properties of the SM
Higgs with two crucial differences: its mass can be in the GeV range and its
couplings to SM matter are suppressed by a universal factor, essentially
$\sin\theta$. The second feature is robust to the extent that
the MSSM decoupling limit can be applied.  As we shall see in
section~\ref{sec:detection}, $\sin\theta$ can be so large that the interactions
of the singlino with the light scalar $\higgs$ lead to a coherent picture of
singlino CDM in which the recent direct detection signals find an explanation.
Before explaining these statements in detail, we will discuss in
section~\ref{sec:Constraints} that the required values of $\sin\theta$ can be
consistent with experimental constraints.

\section{Experimental Constraints on Light Singlets}
\label{sec:Constraints}

We start with a comment on the heavier scalar $h_2$: as mixing
with the light singlets is suppressed, $h_2$ decays like the SM Higgs boson.
Therefore the usual LEP limit $m_{h_2}>114.4\gev$ applies.

Let us now study the light $\higgs$. In experiments the light scalar behaves as
a light SM Higgs with its coupling reduced by the mixing angle $\sin\theta$.
Higgs searches by LEP -- especially the data set from the L3
collaboration~\cite{Acciarri:1996um} -- set strong constraints on the cross
section for $e^+e^-\rightarrow Z + \higgs$ which can be translated directly in
limits on $\sin\theta$.\footnote{Note that in the special case $2 m_{a_s} <
m_\higgs$ the light Higgs can decay into pseudoscalars and the limits get
weaker.} Processes in which the resulting $Z$ decays further into neutrinos are
treated separately.

As we consider values of $m_\higgs$ in the GeV range we also
have to consider the production and subsequent decay of $\higgs$ in meson
decays. The CLEO~\cite{Love:2008hs}, and
BABAR~\cite{Aubert:2009cka,Aubert:2009cp} collaborations have measured the
branching fractions of the radiative decays $\Upsilon\rightarrow \gamma + \ell^+
\ell^-$ with $\ell = \tau,e$. Currently, the limits on $\sin\theta$ from
$\Upsilon$ decay are rather weak (see e.g.~\cite{McKeen:2008gd}), but they may
improve in the future.

\begin{figure}[h!]
\begin{center}
  \includegraphics[width=11.5cm]{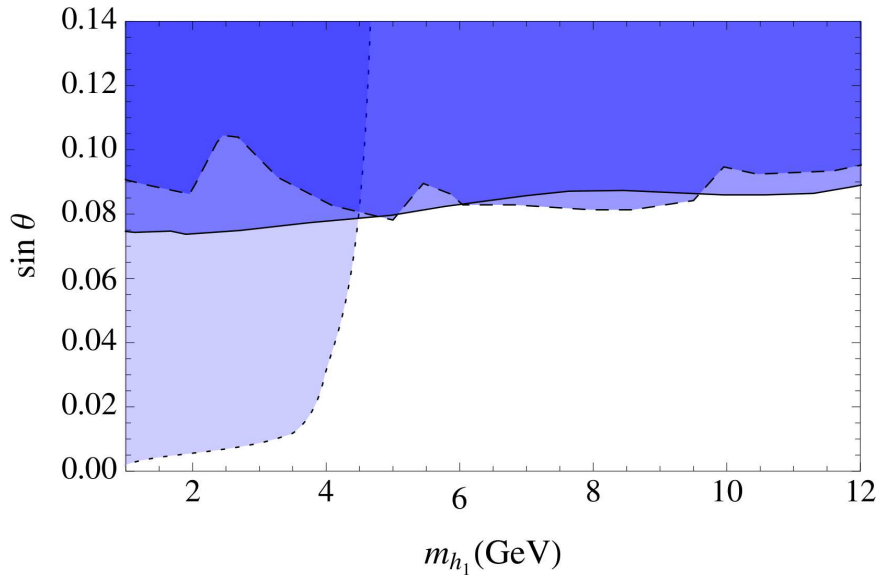}
\end{center}
\caption{Limit on $\sin{\theta}$ from LEP (solid line for neutrino channel,
dashed line for all channels) and B decays (dotted line). The colored  region
is excluded.}
\label{fig:limits}
\end{figure}

Below the $B$ meson threshold $\higgs$ can further contribute to the inclusive
and exclusive decay modes of $B$. Strong limits are set by the inclusive process
$B \rightarrow \higgs + X_s$ followed by the decay $\higgs\rightarrow
\mu^+\mu^-$. The branching ratio for this process can be taken
from~\cite{Alam:1989mta},
\begin{equation}
 \text{Br}(B \rightarrow \higgs + X_s) = 0.058 \left(\frac{\sin\theta}{0.1} \right)^2 \left( 1- \frac{m_\higgs^2}{m_b^2}\right)^2  \,,
\end{equation}
the branching ratio for $\text{Br}(\higgs \rightarrow \mu^+ + \mu^-)$ can be
extracted from~\cite{McKeen:2008gd}.
Measurements of the inclusive $B$ decay by  Belle~\cite{Iwasaki:2005sy} together
with the calculation of the SM background~\cite{Ali:2002jg,Ali:2002ik} suggest
that $ \text{Br}(B \rightarrow \higgs + X_s) \times \text{Br}(\higgs \rightarrow
\mu^+ + \mu^-) < 2.5 \times10^{-6}$. This sets limits on $\sin\theta$ which we
show together with the LEP constraints in figure~\ref{fig:limits}.

\section{Singlinos as Dark Matter}

Let us now discuss whether the singlino discussed above is a
viable dark matter candidate. We start by showing that the singlino has the
right relic abundance to constitute the observed dark matter, continue by
discussing how the interactions with the singlino may explain the current anomalies
in direct detection experiments and finally present a benchmark scenario which
is consistent with present data.

\subsection{Relic Abundance}

\begin{figure}[!ht!]
\centerline{%
\subfigure[\label{fig:annihilation_a}]{\includegraphics{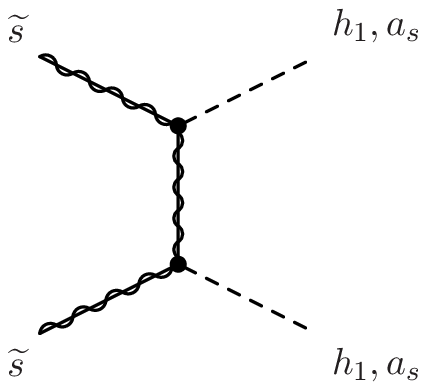}}
\quad
\subfigure[\label{fig:annihilation_b}]{\includegraphics{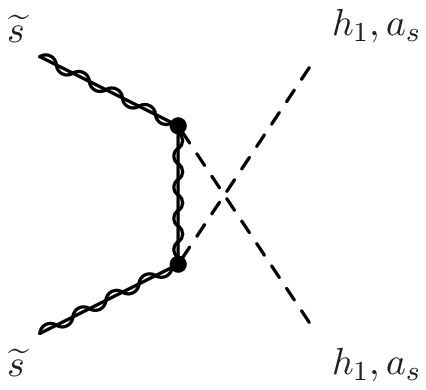}}
\quad
\subfigure[\label{fig:sannihilation_c}]{\includegraphics{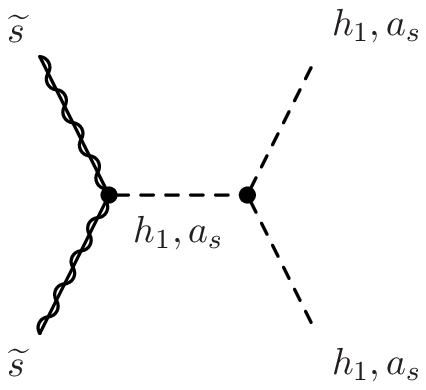}}
}
\caption{Singlino annihilation into (pseudo)scalars.}
\label{fig:AnnihilationDiagrams}
\end{figure}

As singlinos are only very weakly coupled to the MSSM sector,
annihilation into SM particles is suppressed. However, singlinos can efficiently
annihilate into the light singlet scalars\ /\ pseudoscalars provided that
$m_{\singlino}>m_\higgs,m_{\pseudoscalar}$
(see figure~\ref{fig:AnnihilationDiagrams}).
It is convenient to expand the cross section in powers of the relative singlino
velocity $v_\mathrm{rel}$,
\begin{equation}\label{eq:vrelexpansion}
 \sigma\, v_\mathrm{rel} ~=~ \sigma_0 + \sigma_1 \,v_{\mathrm{rel}}^2
 +\mathcal{O}\left( v_{\mathrm{rel}}^4\right) \;.
\end{equation}
As an approximation we will only consider the leading contribution to $\sigma
v_{\mathrm{rel}}$ which is the term $\sigma_0$ for a final state with one scalar
and one pseudoscalar. For final states with either two scalars or two
pseudoscalars $\sigma_0$ vanishes and the term $\sigma_1$ dominates. Evaluating
the Feynman diagrams shown in figure~\ref{fig:AnnihilationDiagrams} we find
the following leading coefficients in the $v_\mathrm{rel}$
expansion~\eqref{eq:vrelexpansion}:
\begin{subequations}
\begin{eqnarray}
 \sigma_1(\singlino\,\singlino~\to~\higgs\,\higgs)
 & \simeq & 
 \frac{17}{256\,\pi}\frac{\kappa^4}{m_{\singlino}^2}\,
 \left(1- \frac{22}{51} \frac{A_\kappa}{m_{\singlino}} +\frac{1}{17}
 \frac{A_\kappa^2}{m_{\singlino}^2}\right)\;,\label{eq:higgshiggs}\\
 \sigma_1(\singlino\,\singlino~\to~\pseudoscalar\,\pseudoscalar)
 & \simeq & 
 \frac{9}{256\,\pi}\frac{\kappa^4}{m_{\singlino}^2}\,\left(1- \frac{14}{27} \frac{A_\kappa}{m_{\singlino}} +\frac{1}{9} \frac{A_\kappa^2}{m_{\singlino}^2}\right)
 \;,\\
 \sigma_0(\singlino\,\singlino~\to~\higgs\,\pseudoscalar)
 & \simeq & 
 \frac{9}{64\,\pi}\frac{\kappa^4}{m_{\singlino}^2}\,
 \left(1+ \frac{2}{3} \frac{A_\kappa}{m_{\singlino}} +\frac{1}{9} \frac{A_\kappa^2}{m_{\singlino}^2}\right)
 \;.
\end{eqnarray}
\end{subequations}
These formulas strictly apply if $m_\higgs,m_{\pseudoscalar}\ll
m_{\singlino}$, but they remain a good approximation as long as $\higgs,\pseudoscalar$ and
$\singlino$ are not degenerate in mass.

The relic singlino density can be obtained from the annihilation cross section
by numerically solving the corresponding Boltzmann equation. An analytic
formula which reproduces our numerical results with good
accuracy is~\cite{Drees:2009bi}
\begin{equation}\label{eq:Omega}
 \Omega_{\widetilde{s}} \,h^2 
 ~=~ 8.5\times 10^{-11}\gev^{-2}\,
 \frac{m_{\widetilde{s}}}{\sqrt{g_*(T_F)}\,T_F\,(\sigma_0 + 3\, T_F\,\sigma_1/m_{\widetilde{s}})}\;,
\end{equation}
where $g_*$ denotes the effective number of relativistic degrees of freedom and
$T_F$ the freeze-out temperature. For reasonable parameter choices we find
$T_F\simeq m_{\widetilde{s}}/20$. Note that in our setup the singlinos typically
freeze out at a temperature close to the QCD phase transition temperature where
the quantity $g_*$ changes rapidly~\cite{Gondolo:1990dk}. This induces an
$\mathcal{O}(1)$ uncertainty in our estimate of the relic density.

Given that the singlet coupling is sizable ($\kappa=\mathcal{O}(0.1)$) and at
least one of the discussed annihilation channels is available, it is possible to
obtain a relic singlino density which matches the observed dark matter density.

\subsection{Direct Detection}\label{sec:detection}

At the same time, singlinos can explain the recently observed
anomalies in direct dark matter detection experiments. The CoGeNT collaboration
has reported an excess of low energy scattering events in their germanium
detector~\cite{Aalseth:2010vx}. This signal is consistent with light weakly
interacting massive particles (WIMPs) ($m\sim 5-15\gev$) which exhibit a rather
large cross section with nucleons, $\sigma_n \sim 10^{-40}\,\text{cm}^2$.
Preliminary data from the CRESST collaboration seem to support this
interpretation~\cite{Seidel:2010nn}, although one should await precise
information on their backgrounds. Particularly strong limits on
light dark matter particles are set by the Xenon10/100
experiment~\cite{Angle:2007uj,Aprile:2010um}. At the moment, however, due to
experimental uncertainties these cannot rule out WIMPs with mass $m \lesssim
10\gev$ as an explanation for CoGeNT and CRESST (see discussion
in~\cite{Collar:2010gg}). A recent analysis~\cite{Hooper:2010uy} suggests that
WIMPs with $m\simeq 7-8\gev$ and $\sigma_n \simeq (1-3) \times
10^{-40}\,\text{cm}^2$ could fit the signals seen at CoGeNT, CRESST and also
DAMA simultaneously. 

\begin{figure}[h!]
\begin{center}
\centerline{%
\includegraphics{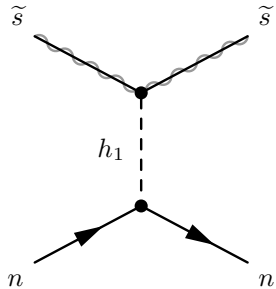}
}
\end{center}
\caption{Singlino nucleon elastic scattering.}
\label{fig:directdetect}
\end{figure}

We will show that the singlino discussed above may have a
direct detection cross section in the range relevant for CoGeNT, CRESST and DAMA.
Apart from its role in giving us the right singlino relic abundance, $\kappa$
enters also the scattering cross section $\sigma_n$ between
singlino CDM and nucleons. This cross section is dominated by
light Higgs exchange (see figure~\ref{fig:directdetect}). The suppression of the
$\higgs$ quark coupling by $\sin{\theta}$ is compensated by the small $m_\higgs$
which enters the denominator of $\sigma_n$ to the forth power. Exchange of
heavier particles like $Z$ or $h_2$ is relatively suppressed, exchange of
$\pseudoscalar$ can be ignored as it is a spin-dependent interaction, and
therefore does not experience the coherent enhancement of the $h_1$ mediated
cross sections.

The cross section for $\higgs$ exchange can be approximated by
\begin{equation}
 \sigma_n~\simeq~\frac{4 \,m_{\singlino}^2\, m_n^2}{\pi\,(m_{\singlino}+m_n)^2}\,
 f_n^2~\simeq~\frac{4 \, m_n^2}{\pi}\,
 f_n^2\;.
\end{equation}
Here $m_n$ denotes the nucleon mass and $f_n$ the effective singlino nucleon
coupling which can be expressed as
\begin{equation}
 f_n~=~m_n\,f_q\,\left(f^n_u + f^n_d + f^n_s + \frac{6}{27} f^n_G \right)\,,
\end{equation}
where $f_q$ is the singlino quark coupling divided by the quark mass. In our model we have
\begin{equation}
 f_q~=~g_{\higgs \singlino\singlino}\; \frac{\sin{\theta}}{\sqrt{2} \, v_\mathrm{EW}}\,\frac{1}{m_\higgs^2}\,.
\end{equation}
Furthermore, $f^n_u$, $f^n_d$, $f^n_s$ and $f^n_G$ specify the up-, down-,
strange-quark and gluon contribution to the nucleon mass which were determined
in pion-nucleon scattering experiments. The cross section
from~\cite{Pavan:2001wz} translates into $f^n_u\simeq0.03$, $f^n_d\simeq0.04$,
$f^n_s\simeq0.38$ and $f^n_G\simeq0.55$. Note, however, that these quantities are
subject to large uncertainties. 
Numerically we find
\begin{equation}\label{eq:SigmaNeutron}
 \sigma_n ~\sim~ 10^{-40}\,\text{cm}^{2}\,\left( \frac{\kappa}{0.08}\right)^2\left(\frac{\sin{\theta}}{0.03} \right)^2\,\left( \frac{4\gev}{m_\higgs}\right)^4\;.
\end{equation}

\subsection{A Benchmark Scenario}

\begin{table}[!ht!]
\centerline{%
\subtable[Input parameters.]{%
\begin{tabular}{ccc}
\hline\hline
Quantity &$\quad$& Value\\
\hline
$\mu_\text{eff}$&& $370\gev$\\
$x$&& $163\gev$\\
$A_{\kappa}$&&$-9\gev$\\
$\mu_s$&&$-19\gev$\\
$B\mu_s$&&$0$\\
$\lambda$&&$-0.003$\\
$\kappa$&&$0.08$\\
\hline\hline
\end{tabular}
}
\qquad
\subtable[Predictions.]{%
\begin{tabular}{ccc}
\hline\hline
Quantity &$\quad$& Value\\
\hline
$m_{\pseudoscalar}$&&$28\gev$\\
$m_{\tilde{s}}$&&$7\gev$\\
$m_{h_1}$&& $4\gev$\\
$\sin\theta$&& $0.03$\\
$\sigma_n$&& $\sim 10^{-40}\,\text{cm}^2$\\
$\Omega\, h^2$&& $\sim 0.1$\\
\hline\hline
\end{tabular}
}
}
\caption{Parameters of a phenomenologically viable benchmark point. We assume
$m_{h}=115\gev$.}
\label{tab:benchmark}
\end{table}

To illustrate our results, let us look at some benchmark values
(table~\ref{tab:benchmark}). From \eqref{eq:Omega} it follows that the relic
abundance of the singlino LSP has the appropriate value to match the observed CDM
density. Due to the mass relation $m_{a_s}>m_\singlino > m_\higgs$
annihilation can only proceed into $\higgs\higgs$ final states and the cross
section is determined by~\eqref{eq:higgshiggs}.
Equation~\eqref{eq:SigmaNeutron} shows that the singlino nucleon cross section
is in a range where the CoGeNT, CRESST and DAMA signals can potentially be
explained.

\section{Conclusion}
\label{sec:Conclusions}

We have discussed a simple singlet extension of the MSSM in which the singlino
LSP can constitute the observed cold dark matter of the universe. There is a
scalar particle $\higgs$ with mass in the few GeV region which behaves like the
SM Higgs with universally reduced couplings.  We have checked that the light
$\higgs$ is consistent with collider and flavor physics constraints. An
important ingredient of our scenario is the $\higgs$ singlino coupling $\kappa$,
which is of the order 0.1. This facilitates efficient annihilation of singlinos
into $\higgs$ pairs, which decay further into quark and lepton pairs, such that
the correct relic abundance can be obtained. The same coupling $\kappa$ enters
$\higgs$ mediated interactions with nuclei, which can potentially explain the
CoGeNT, CRESST and DAMA anomalies. 

Our scenario will soon be tested in various experiments. Future direct detection
experiments will confirm or rule out the dark matter interpretation of CoGeNT,
CRESST and DAMA. Neutrino telescopes will soon reach the sensitivity where they
can probe singlino annihilation in the sun, especially if a significant fraction
of the annihilation products are taus. The hypothesis of a singlino LSP can be
tested  at the LHC. Promising signatures include the measurement of missing
energy which is reduced against what one expects in the usual neutralino case.
Further, the next-to-lightest superpartner may be charged, which can result in
charged tracks and other interesting signatures.  Finally, $B$ factories offer
the possibility to look for the light scalar $\higgs$ in decays of $\Upsilon$
and $B$ mesons.

\subsection*{Acknowledgments} 
We would like to thank Andrzej Buras for discussions and Kai Schmidt-Hoberg for
collaboration in the early stages of the project. This research was supported by
the DFG cluster of excellence Origin and Structure of the Universe, and the
\mbox{SFB-Transregio} 27 "Neutrinos and Beyond" by Deutsche
Forschungsgemeinschaft (DFG).

\providecommand{\bysame}{\leavevmode\hbox to3em{\hrulefill}\thinspace}


\begin{thebibliography}{10}

\bibitem{Aalseth:2010vx}
CoGeNT, C.E. Aalseth et~al.,  1002.4703.
%%CITATION = 1002.4703;%%

\bibitem{Seidel:2010nn}
T. by~W.~Seidel,  WONDER 2010 Workshop, Gran Sasso, Italy, March 22-23, 2010
  and IDM 2010, University of Montpellier, France, July 26-30, 2010.

\bibitem{Bernabei:2008yi}
DAMA, R. Bernabei et~al., Eur. Phys. J. \textbf{C56} (2008), 333--355.
%%CITATION = 0804.2741;%%

\bibitem{Bernabei:2010mq}
R. Bernabei et~al., Eur. Phys. J. \textbf{C67} (2010), 39--49.
%%CITATION = 1002.1028;%%

\bibitem{Hooper:2002nq}
D. Hooper and T. Plehn, Phys. Lett. \textbf{B562} (2003), 18--27.
%%CITATION = HEP-PH/0212226;%%

\bibitem{Bottino:2002ry}
A. Bottino, N. Fornengo and S. Scopel, Phys. Rev. \textbf{D67} (2003), 063519.
%%CITATION = HEP-PH/0212379;%%

\bibitem{Feldman:2010ke}
D. Feldman, Z. Liu and P. Nath, Phys. Rev. \textbf{D81} (2010), 117701.
%%CITATION = 1003.0437;%%

\bibitem{Das:2010ww}
D. Das and U. Ellwanger, JHEP \textbf{09} (2010), 085.
%%CITATION = 1007.1151;%%

\bibitem{Hooper:2009gm}
D. Hooper and T.M.P. Tait, Phys. Rev. \textbf{D80} (2009), 055028.
%%CITATION = 0906.0362;%%

\bibitem{Draper:2010ew}
P. Draper et~al.,  1009.3963.
%%CITATION = 1009.3963;%%

\bibitem{Belikov:2010yi}
A.V. Belikov et~al.,  1009.0549.
%%CITATION = 1009.0549;%%

\bibitem{Drees:1988fc}
M. Drees, Int. J. Mod. Phys. \textbf{A4} (1989), 3635.
%%CITATION = IMPAE,A4,3635;%%

\bibitem{Kim:1983dt}
J.E. Kim and H.P. Nilles, Phys. Lett. \textbf{B138} (1984), 150.
%%CITATION = PHLTA,B138,150;%%

\bibitem{Giudice:1988yz}
G.F. Giudice and A. Masiero, Phys. Lett. \textbf{B206} (1988), 480--484.
%%CITATION = PHLTA,B206,480;%%

\bibitem{Delgado:2010cw}
A. Delgado et~al.,  1005.4901.
%%CITATION = 1005.4901;%%

\bibitem{LRRRSSV2}
H.M. Lee et~al., {\emph{in preparation}}.

\bibitem{Acciarri:1996um}
L3, M. Acciarri et~al., Phys. Lett. \textbf{B385} (1996), 454--470.
%%CITATION = PHLTA,B385,454;%%

\bibitem{Love:2008hs}
CLEO, W. Love et~al., Phys. Rev. Lett. \textbf{101} (2008), 151802.
%%CITATION = 0807.1427;%%

\bibitem{Aubert:2009cka}
BABAR, B. Aubert et~al., Phys. Rev. Lett. \textbf{103} (2009), 181801.
%%CITATION = 0906.2219;%%

\bibitem{Aubert:2009cp}
BABAR, B. Aubert et~al., Phys. Rev. Lett. \textbf{103} (2009), 081803.
%%CITATION = 0905.4539;%%

\bibitem{McKeen:2008gd}
D. McKeen, Phys. Rev. \textbf{D79} (2009), 015007.
%%CITATION = 0809.4787;%%

\bibitem{Alam:1989mta}
CLEO, M.S. Alam et~al., Phys. Rev. \textbf{D40} (1989), 712--720.
%%CITATION = PHRVA,D40,712;%%

\bibitem{Iwasaki:2005sy}
Belle, M. Iwasaki et~al., Phys. Rev. \textbf{D72} (2005), 092005.
%%CITATION = HEP-EX/0503044;%%

\bibitem{Ali:2002jg}
A. Ali et~al., Phys. Rev. \textbf{D66} (2002), 034002.
%%CITATION = HEP-PH/0112300;%%

\bibitem{Ali:2002ik}
A. Ali,  hep-ph/0210183.
%%CITATION = HEP-PH/0210183;%%

\bibitem{Drees:2009bi}
M. Drees, M. Kakizaki and S. Kulkarni, Phys. Rev. \textbf{D80} (2009), 043505.
%%CITATION = 0904.3046;%%

\bibitem{Gondolo:1990dk}
P. Gondolo and G. Gelmini, Nucl. Phys. \textbf{B360} (1991), 145--179.
%%CITATION = NUPHA,B360,145;%%

\bibitem{Angle:2007uj}
XENON, J. Angle et~al., Phys. Rev. Lett. \textbf{100} (2008), 021303.
%%CITATION = 0706.0039;%%

\bibitem{Aprile:2010um}
XENON100, E. Aprile et~al., Phys. Rev. Lett. \textbf{105} (2010), 131302.
%%CITATION = 1005.0380;%%

\bibitem{Collar:2010gg}
J.I. Collar and D.N. McKinsey,  1005.0838.
%%CITATION = 1005.0838;%%

\bibitem{Hooper:2010uy}
D. Hooper et~al.,  1007.1005.
%%CITATION = 1007.1005;%%

\bibitem{Pavan:2001wz}
M.M. Pavan et~al., PiN Newslett. \textbf{16} (2002), 110--115.
%%CITATION = HEP-PH/0111066;%%

\end{thebibliography}
\end{document}